\documentclass[%
 reprint,
superscriptaddress,
 amsmath,amssymb,
 aps,
pra,
]{revtex4-1}

\usepackage{latexsym,amsmath,amssymb}
\usepackage{braket} 
\usepackage{ftnxtra}
\usepackage{fnpos}
\usepackage{tikz}
\usepackage{color}
\usepackage{footnote}

\newcommand{\nn}{\nonumber}

\usepackage{amsthm} 
\newtheorem{theorem}{Theorem}

\newtheorem{definition}{Definition}[theorem]
\newtheorem{result}{Result}

\usepackage[T1]{fontenc} 

\usepackage{graphicx}
\usepackage{epstopdf}

\newcommand{\ba}{\begin{eqnarray}}
\newcommand{\ea}{\end{eqnarray}}
\newcommand{\ban}{\begin{eqnarray*}}
\newcommand{\ean}{\end{eqnarray*}}

\begin{document}
\title{Self-testing of symmetric three-qubit states}
\date{\today}
\author{Xinhui Li}
 \affiliation{State Key Laboratory of Networking and Switching Technology,
Beijing University of Posts and Telecommunications, Beijing, China 100876}

\author{Yukun Wang}
 \email{wykun06@gmail.com}
 \affiliation{ Department of Electrical \& Computer Engineering,
National University of Singapore, Singapore, 117543}

\author{Yunguang Han}
 \email{hyg@fudan.edu.cn}
 \affiliation{Department of Physics and Center for Field Theory and Particle Physics,\\
Fudan University, Shanghai, China, 200433}
 \affiliation{State Key Laboratory of Surface Physics, 
Fudan University, Shanghai, China 200433}

\author{Fei Gao}
 \email{Gaof@bupt.edu.cn}
 \affiliation{State Key Laboratory of Networking and Switching Technology,
Beijing University of Posts and Telecommunications, Beijing, China 100876}
\affiliation{Center for Quantum Computing, Peng Cheng Laboratory,
Shenzhen, China, 518055}

\author{Qiaoyan Wen}
 \affiliation{State Key Laboratory of Networking and Switching Technology,
Beijing University of Posts and Telecommunications, Beijing, China 100876}

\begin{abstract}
Self-testing refers to a device-independent way to uniquely identify the state 
and the measurement for uncharacterized quantum devices. 
The only information required comprises the number of 
measurements, the number of outputs of each measurement, 
and the statistics of each measurement. Earlier results on self-testing of
multipartite state were restricted either
to Dicke states or graph states. In this paper, we propose self-testing schemes for a large family of symmetric three-qubit states, namely the superposition of $W$ state and $GHZ$ state. We first propose and analytically prove a self-testing criterion for the special symmetric state with equal coefficients of the canonical basis, by designing subsystem self-testing of partially and maximally entangled state simultaneously. Then we demonstrate for the general case,  the states can be self-tested numerically 
by the swap method combining semi-definite programming (SDP) in high precision.

\end{abstract}

\maketitle


\section{Introduction}
\label{sec:introduction}

Entanglement is a critical resource 
 for numerous striking applications of quantum information theory
\cite{Horodecki}.  Furthermore, it is key to comprehend many peculiar
properties of quantum many-body systems and
has become increasingly important in both theoretical and experimental
areas such as teleportation \cite{Bennett} and quantum simulation \cite{Reichardt}.
Because of the essential role played by symmetry in the field of quantum entanglement, it is of great significance to explore the properties of symmetric states. Also symmetric states are key resources in many experiments such as quantum communication\cite{Brandao}, and quantum computing, for instance as an initial state for Grover's algorithm \cite{Ivanov}. In addition, restricting analysis to symmetric states can greatly reduce the difficulty of calculations. In this work, we investigate one of verification tasks for symmetric states, namely certification of entanglement state.

A canonical way to approach the problem of certification of quantum states is
to exploit tomographic scenario \cite{Kosaka}. By repeating the experiment,
expectation values of an informationally complete
set of measurements allow us to reconstruct the density
operator that describes the quantum state. However, 
such procedure requires a large number fully characterized measurements 
that scales with the dimension of the quantum state.

An alternative technique which could positively address these problems is \textit{self-testing}.
It is a concept of device independence whose conclusion verdict relies only on the observed statistics of measurement outcomes under the sole assumptions of no-signaling and the validity of quantum theory\cite{Scarani}. Consider two players, Alice and Bob, each has a device. Both devices are given classical input ($x$ and $y$, respectively) which corresponds to the application of measurements inside the devices, and classical output ($a$ and $b$). The devices are physical isolated so that sending signals from one to the other is not possible. The central question is: given observed correlational probabilities $p(a,b|xy)$ ,
what can be inferred about the underlying state? 
Self-testing refers to determining the state completely in such cases.

The idea of self-testing quantum states  can be traced back to 1990's, 
where Popescu and Rohrlich et al. pointed out the maximal violation 
of the CHSH Bell inequality \cite{Clauser} identifies uniquely the maximally 
entangled state of two qubits and the corresponding measurements \cite{Tsirelson}. 
However, it was not widely known until the works of Mayers and Yao \cite{Mayers}, 
which self-tests the same state with more measurements. 
Since then self-testing has received substantial attention: self-testing
for partially entangled pairs of qubits were presented in \cite{Yang,Bamps}, 
while its extension to high dimension partially entangled states was given in \cite{Coladangelo}. 
Furthermore, all the criteria for self-testing the maximally
entangled pair of qubits were reported in refs \cite{Miller,Wang}, where
the authors proved a condition for a given binary XOR game to
be a robust self-test. The robustness analysis to small 
deviations from the idea case for self-testing these quantum states
and measurements were presented in \cite{McKague,Kaniewski,Bancal,Yang2}, 
which made self-testing more practical.

Beyond these works focusing on the bipartite scenario, self-testing of 
multipartite states have recently been studied, such as self-testing of Graph states \cite{McKague2}, 
Dicke states \cite{Supic}, partially entangled GHZ states \cite{Supic}. Inspired by self-testing 
all entangled states in bipartite scenario, one may ask whether all the entanglement
states can be self-tested in multipartite scenario? However, the multipartite entanglement 
is more complicated than bipartite scenario, especially for partially 
entangled states. The most celebrated example is the case of three-qubit states \cite{Acin1,Acin2}, 
we consider the particular case of the sates which are equivalent under local unitary transformations to states of the form\cite{Linden},\[\ket{\eta}=a\ket{000}+b\ket{001}+c\ket{010}+d\ket{100}+e\ket{111}\]
where $a,b,c,d,e$ are normalized coefficients. The two well known inequivalent classes of tripartite genuine entangled states, namely, $W$ states \cite{DurW} and Greenberger-Horne-Zeilinger ($GHZ$) states \cite{Brunner} are corresponding to $\frac{1}{\sqrt{3}}(\ket{001}+\ket{010}+\ket{100})$ and $\frac{1}{\sqrt{2}}(\ket{000}+\ket{111})$ respectively. 
It is obviously that entangled three-qubit states are not only these two kinds of entangled states. One can set any value of the coefficients to define entanglement states. However, what we're interested in is the symmetric entangled states, due to its significant application. We noticed that the symmetric entangled three-qubits states are of the form $\cos\theta\ket{W}+\sin\theta\ket{GHZ}$ under permutations of party labels. The aim of this paper is to investigate the self-testing of these states and where so far only special case has been studied, i.e.,  self-testing 
of $W$ state \cite{Wu} and self-testing of $GHZ$ state \cite{Kaniewski}.

Here we proved analytically the self-testing of a specific symmetric state through projections onto two systems and showed that general cases can be self-tested using fixed Pauli measurements combining the swap method and semidefinite programming (SDP). The paper is structured as following. In Section \ref{sec:pre},
we give a review of the two-qubit self-testing, including the whole set of criteria for ideal
self-testing of maximally entanglement state and any pure two-qubit state can be self-tested by tilted Bell inequality. In Section \ref{sec:main work}, we prove that a  symmetric three-qubit state can be self-tested through projections onto two systems, and we show robust self-testing of a more general class of
states which is a linear combination of $W$ and $GHZ$ states by the swap method and SDP.

\section{Preliminaries}\label{sec:pre}

Let us consider a Bell-type experiment involving two noncommunicating parties. Each has access to a black box with inputs denoted respectively by $x,y\in \left\{0,1,...,M-1\right\}$ and outputs $a,b\in \left\{0,1,...,m-1\right\}$. Assuming the validity of quantum mechanics, one could model these boxes with an underlying state $\ket{\psi}_{AB}$ and measurement projectors $\left\{M_x^a\right\}_{x,a}$ and $\left\{M_y^b\right\}_{y,b}$, which commute for different parties. The state can be taken pure and the measurements can be taken projective without loss of generality, because the dimension of the Hilbert space is not fixed and the possible purification and auxiliary systems can be given to any of the parties. After sufficiently many repetitions of the experiment one can estimate the joint conditional statistics, also known as \emph{the behavior}, $p(a,b| x,y) = \bra{\psi} M_x^aM_y^b \ket{\psi}$. Now, we can formally define self-testing in the following way.
\begin{definition}
(Self-testing) We say that the correlations $p(a,b|x,y)$ allow for self-testing if for every quantum behavior
$(\ket{\psi}$, $\{M_x^a,M_y^b\})$ compatible with $p(a,b|x,y)$ there exists a local isometry $\Phi=\Phi_A\otimes\Phi_B$ such that
\begin{align}\label{eq:self-testing definiton}
&\Phi\ket{\psi}_{AB}\ket{00}_{A'B'}
=\ket{junk}_{AB}\otimes\ket{\overline{\psi}}_{A'B'}\\
&\Phi(M^a_x \otimes M^b_y\ket{\psi}_{AB}\ket{00}_{A'B'})
=\ket{junk}_{AB}\otimes \overline{M^a_x}\otimes\overline{M^b_y}\ket{\overline{\psi}}_{A'B'},\nn
\end{align}
where $|00\rangle_{A'B'}$ is the trusted auxiliary qubits attached by Alice and
Bob locally into their systems\cite{Supic1}. The isometry must be seen as a virtual protocol: 
it does not need to be implemented in the laboratory as a part of the procedure 
of self-testing; all that must be done in laboratory is to query the boxes and 
derive $p(a,b|x,y)$.
\end{definition}

Let us review some previous results on the self-testing of 
two-qubit state which are used as building blocks of our work.

\subsection{All the self-testings of the singlet for two binary
measurements}

In the ref.\cite{Wang}, Wang et al. proposed the whole set of criteria for the ideal self-testing of singlet.
Consider four unknown operators $A_x$ and $B_y$ for $i,j\in\{1,2\}$ with binary outcomes labelled $\pm1$ and satisfy $[A_x, B_y]=0$. Denote
\begin{equation}\label{eq:ang}E_{xy}=\langle \psi|A_x B_y |\psi\rangle=\cos\alpha_{xy}\end{equation}
where $\alpha_{xy}$ is the angle between the two vectors $A_x|\psi\rangle$ and $B_y|\psi\rangle$. The observed correlations
$E_{xy}$
self-test the singlet if and only if they satisfy one of the conditions
\begin{equation}
\label{t2}
\sum_{(x,y)\neq(i,j)}\arcsin (E_{xy})-\arcsin(E_{i,j})=\xi\pi,
\end{equation}
  with $\arcsin(E_{xy})_{\{x,y\}\in\{1,2\}}\in[-\frac{\pi}{2}, \frac{\pi}{2}]$ for $i,j\in\{1,2\}$, $\xi\in\{+1,-1\}$. The eight equations in \eqref{t2} are equivalent in the sense that each one can be transformed into the other by relabelling the measurements and outcomes. Without loss of generality, consider the case of $i=1$, $j=2$ and $\xi=+1$, that is $\alpha_{11}+\alpha_{21}=\alpha_{12}-\alpha_{22}$.
It means $A_1|\psi\rangle$, $A_2|\psi\rangle$, $B_1|\psi\rangle$, $B_2|\psi\rangle$ are in the same plane.

This all self-testing criteria for singlet state is proved to be equivalent to a binary nonlocal XOR game defined by the figure of merit $\sum\limits_{(x,y)\in\{1,2\}^2}f_{xy}E_{xy}$ if $E_{xy}=\cos \alpha_{xy}$ satisfy $\alpha_{11}+\alpha_{21}=\alpha_{12}-\alpha_{22}$ and the coefficients $f_{xy}$ are constructed by
 \begin{equation}\label{eq:XOR}
 \left(\begin{matrix}f_{11}\\
  f_{12}\\
  f_{21}\\
  f_{22}\end{matrix}\right)=\left(\begin{matrix}\frac{1}{\sin \alpha_{11}}\\
  -\frac{1}{\sin(\alpha_{11}+\alpha_{21}+\alpha_{22})}\\
  \frac{1}{\sin\alpha_{21}}\\
  \frac{1}{\sin\alpha_{22}}\end{matrix}\right).
 \end{equation}

\subsection{Self-testing of pure partially entangled two-qubit state }\label{sec:2P}

It has been shown that any pure two-qubit state
in their Schmidt form
\begin{align}
\ket{\psi}=\cos\theta\ket{00}+\sin\theta\ket{11}
\end{align}
can be self-tested by observing the maximum violation of the
tilted CHSH inequality \cite{Yang,Bamps,Bancal}
\begin{align}
I^\alpha(\alpha,A_1,A_2,B_1,B_2)=&\alpha A_1+A_1B_1+A_1B_2\nn\\
&+A_2B_1-A_2B_2\leq 2+\alpha
\end{align}
where $\alpha$ is defined through $\sin2\theta=\sqrt{\frac{4+\alpha^2}{4-\alpha^2}}$.
$A_1$, $A_2$, $B_1$ and $B_2$ are the unknown
measurements by Alice and Bob, respectively.
The maximal quantum violation of this inequality is given
by $I^\alpha_Q=\sqrt{8+2\alpha^2}$, achievable with the measurement
settings
\begin{align}\label{M Tilted}
A_1&=\sigma_z,\;\;\;A_2=\sigma_x;\nn \\
B_1&=\cos\mu \sigma_z+\sin\mu\sigma_x,\nn \\
B_2&=\cos\mu \sigma_z-\sin\mu\sigma_x,
\end{align}
where $\tan\mu=\sin2\theta$.

\section{Self-testing of symmetric three-qubit states}\label{sec:main work}

The work in \cite{Acin1,Acin2} gave a generalization of the Schmidt decomposition for three-qubit
pure states and proved that for any pure three-qubit state the existence
of local bases which allow one to build a set of five
orthogonal product states in terms of which can be
written in a unique form. The local bases product states can be given as
three inequivalent sets
\begin{align}\label{Sta:Acin}
\{\ket{000},\;\ket{001},\;\ket{010},\;\ket{100},\ket{111}\},\nn\\
\{\ket{000},\;\ket{010},\;\ket{110},\;\ket{100},\ket{111}\},\nn\\
\{\ket{000},\;\ket{100},\;\ket{110},\;\ket{101},\ket{111}\},
\end{align}
whereas the first set is symmetric under permutation of
parties, the other two are not.

In this paper, we consider the self-testing of symmetric
three-qubit states based on the fist set of \eqref{Sta:Acin}
with different kinds of coefficients.

\subsection{Self-testing of a symmetric three-qubit state}\label{sec:3.1}

\begin{figure}[ht]\label{fig:SWAP}
\centering
\includegraphics[scale=0.6]{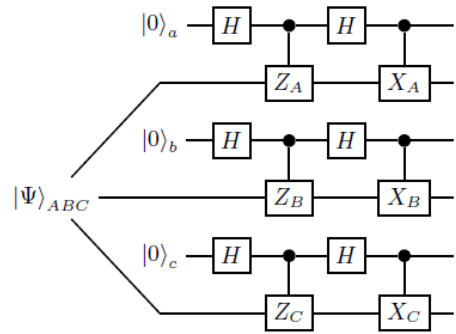}
\caption{The swap circuit. The local isometry used 
to self-test the $\ket{\psi}$ state. 
$H$ is the standard Hadamard gate, 
$Z$ and $X$ are controlled by the auxiliary qubit. 
The trusted ancillary qubits are prepared in the state $\ket{0}$.}
\end{figure}

The specific case we consider is the state with equal coefficient of the basis, reads as:
\begin{equation}\label{eq:target1}
\ket{\psi}=\frac{1}{\sqrt{5}}(\ket{001}+\ket{010}+\ket{100}+\ket{000}+\ket{111}).
\end{equation}

The basic idea of self testing this state is to project the state  onto  two kinds of subsystem entangled states by one party's measurement $Z$. More precisely, after measuring one party in the $Z$ basis, the remaining two parties can achieve the maximal violation of  tailored Bell inequalities simultaneously using the same measurement settings conditioned on the outcome being either "0" or "1".  

If partition the three parties into $A|BC$, we have
\begin{align}\label{Sta:A|BC}
\ket{\psi}=&\frac{1}{\sqrt{5}}[\sqrt{3}\ket{0}_{A}
\otimes\frac{1}{\sqrt{3}}(\ket{00}+\ket{01}+\ket{10})_{BC}\nn\\
&+\sqrt{2}\ket{1}_A\otimes\frac{1}{\sqrt{2}}(\ket{00}+\ket{11})_{BC}].
\end{align}
We denote
   $$\ket{\varphi_0}:=\frac{1}{\sqrt{3}}(\ket{00}+\ket{01}+\ket{10}).$$
Using Schmidt decomposition, the state $\ket{\varphi_0}$ can be written as
   \begin{align}
   \ket{\varphi_0}=\cos\beta\ket{0'0'}+\sin\beta\ket{1'1'},
   \end{align}
here $\cos \beta =\sqrt{\frac{3+\sqrt{5}}{6}}$, $\sin\beta=\sqrt{\frac{3-\sqrt{5}}{6}}$ and $\{\ket{0'},\ket{1'}\}$ are the new basis (see the detail in Appendix). 
Following the results given in sec \ref{sec:2P}, this state can be self-tested 
by violating  the tilted CHSH inequality maximally
\begin{align}I_Q^\alpha=\sqrt{8+2\alpha^2}=\frac{12}{\sqrt{13}},\end{align}
with $\alpha=2\sqrt{\frac{1-\sin^22\beta}{1+\sin^22\beta}}$.
The optimal measurements are set according to \eqref{M Tilted}, with $\tan\mu =\frac{1}{3}$.

At the same time, singlet is invariant under basis transformation
\begin{align}
\ket{\varphi_1}=\frac{\ket{0'0'}+\ket{1'1'}}{\sqrt{2}}
\end{align}
using the same bases and optimal measurements settings with $\ket{\varphi_0}$
would satisfy $\alpha_{11}+\alpha_{21}=\alpha_{12}-\alpha_{22}$ where $\alpha_{11}=\mu,\;\alpha_{12}=-\mu,\;\alpha_{21}=\frac{\pi}{2}-\mu$ and $\alpha_{22}=-\frac{\pi}{2}-\mu$.
So, $\ket{\varphi_1}$ can be self-tested by
winning the binary nonlocal XOR game \cite{Wang}:
\begin{align} \sum\limits_{(x,y)\in\{1,2\}^2}f_{xy}\cos\alpha_{xy}=\frac{4}{\sin2\mu}\end{align}
and the coefficients $f_{xy}$ are constructed as \eqref{eq:XOR}.

Hence, the states  $\ket{\varphi_0}$ and $\ket{\varphi_1}$ 
conditioned on the outcome "0" and "1" after the measurement in the $Z$ basis of $A$ violate the tilted CHSH inequality and XOR game maximally using the same measurements, respectively. This also holds when the first measured party is $C$.

The following result sums it up.

\begin{result}
Alice, Bob, and Charlie, spatially separated,
each perform three measurements denoted as $\{A_i,B_j,C_k\}$ ($i,j,k\in\{0,1,2\}$) 
with binary outcomes on an
unknown shared quantum state $\ket{\psi}$. The target state is self-tested
if the following statistics are observed:
\begin{align}\label{eq:Cor1}
\langle P^0_AP^0_BP^0_C\rangle &=\langle P^0_AP^0_BP^1_C\rangle=\langle P^0_AP^1_BP^0_C\rangle\nn \\
&=\langle P^1_AP^0_BP^0_C\rangle=\langle P^1_AP^1_BP^1_C\rangle=\frac{1}{5},
\end{align}
\begin{subequations}
\begin{align}\label{ZAperXB}
\langle P^i_CA_1(B_2-B_1)\rangle&=0,\\
\label{ZAXA}
\cos \omega \langle P^i_CA_1\rangle+\sin\omega \langle P^i_CA_2\rangle&=\langle P^i_C A_0\rangle,
\end{align}
\end{subequations}
\begin{subequations}
\begin{align}
\langle P^i_A (B_1-B_2)C_1\rangle&=0,\\
\cos \omega \langle P^i_A C_1\rangle+\sin\omega \langle P^i_AC_2\rangle&=\langle P^i_AC_0\rangle,
\end{align}
\end{subequations}
\begin{subequations}
\begin{align}
\langle P^i_C A_1(B_1-B_2)\rangle&=0,\\
\frac{\sin(\omega+\mu)\langle P^i_CB_2\rangle}{\sin2\mu}
-\frac{\sin(\omega-\mu)\langle P^i_CB_1\rangle}{\sin2\mu}&=\langle P^i_CB_0\rangle
\end{align}
\end{subequations}
for $i\in\{0,1\}$, and
\begin{subequations}
\begin{align}\label{Cor:XCXA}
 \sin \omega \langle P^0_xP^0_y T_2\rangle-\cos\omega \langle P^0_xP^0_yT_1\rangle&=\frac{2}{5}, \\
 \label{Cor:XB}
  \frac{\cos(\omega-\mu)\langle P^0_AP^0_CB_1\rangle}{\sin2\mu}-
    \frac{\cos(\omega+\mu)\langle P^0_AP^0_CB_2\rangle}{\sin2\mu}&=\frac{2}{5}
    \end{align}
\end{subequations}
for $(x,y,T)=\{(A,B,C),(B,C,A)\}$, and
\begin{subequations}
\begin{align}\label{eq:Tilted}
\langle P_C^0 I^\alpha(\alpha, A_1, A_2,B_1,B_2) \rangle &=\frac{3}{5}\sqrt{8+2\alpha^2}, \\
\langle P_A^0 I^\alpha(\alpha, C_1, C_2,B_1,B_2)\rangle &=\frac{3}{5}\sqrt{8+2\alpha^2},
\end{align}
\end{subequations}
\begin{align}\label{eq:CHSH}
 \sum \limits_{{i,j}=\{1,2\}}f_{i,j}\langle P_A^1 B_iC_j\rangle=\sum\limits_{{i,j}=\{1,2\}} f_{ij}\langle P_C^1 A_iB_j\rangle &=\frac{8}{5\sin2\mu}
\end{align}
where $P^0=\frac{1+Z}{2}$ and $P^1=\frac{1-Z}{2}$ 
are projectors for the $Z$ measurement, $\alpha=2\sqrt{\frac{5}{13}}$,
$\tan\mu=\frac{2}{3}$, $\cos\omega=\frac{1}{\sqrt{5}}$ and $\sin\omega=\frac{2}{\sqrt{5}}$.
\end{result}

\textbf{Proof.} We start from observation \eqref{eq:Cor1} which implies that
\begin{align}
\langle P^0_AP^0_BP^0_C\rangle &+\langle P^0_AP^0_BP^1_C\rangle +\langle P^0_AP^1_BP^0_C\rangle\nn \\
&+\langle P^1_AP^0_BP^0_C\rangle+\langle P^1_AP^1_BP^1_C\rangle=1.
\end{align}
Therefore $P_A^aP_B^bP^c_C\ket{\psi}=0$ for other three projectors.

For convenience, define the operators for each party as
\begin{align} 
Z_A=&A_0,\;\; \;\;Z'_A=A_1,\;\; \;\;X'_A=A_2; \nn\\
Z_B=&B_0,\;\; \;\;Z'_B=\frac{B_1+B_2}{2\cos\mu},\;\;\;\;X'_B=\frac{B_1-B_2}{2\sin\mu};\nn\\
Z_C=&C_0,\;\; \;\;Z'_C=C_1,\;\;\;\;X'_C=C_2.
\end{align}

Following the self-testing of nonmaximally entangled
qubits from the ref. \cite{Bamps}, maximal violation of the tilted Bell inequality \eqref{eq:Tilted}
implies
\begin{subequations}\label{Til}
\begin{align}
\label{relation:Til1}
P^0_CZ'_A\ket{\psi}&=P^0_CZ'_B\ket{\psi},  \\
\label{relation:Til2}
P^0_CZ'_AX'_A\ket{\psi}&=-P^0_CX_A'Z'_A\ket{\psi},\\
\label{relation:Til3}
p^0_CX'_A(I+Z_B')\ket{\psi}&=\frac{1}{\tan\theta} P^0_C X'_B(1-Z'_A)\ket{\psi}.
\end{align}
\end{subequations}
From \eqref{relation:Til1} and \eqref{relation:Til3}, we get 
\begin{subequations}
\begin{align}
\label{relation:Til3.3}
P^0_C X'_B(1+Z'_A)\ket{\psi}=&\frac{1}{\tan\theta}P^0_C X'_A(1-Z'_B)\ket{\psi}, \\
\label{relation:Til3.2}
P^0_C X'_B(1-Z'_B)\ket{\psi}=&\tan\theta P^0_C X'_A(1+Z'_A)\ket{\psi}.   
\end{align}
\end{subequations}
On the other hand, the following equation holds with \eqref{relation:Til1},
\begin{align}\label{relation:Til3.1}
P^0_C X'_B\ket{\psi}=&P^0_C X'_B\frac{(1+Z'_A)}{2}\ket{\psi}+P^0_C X'_B\frac{(1-Z'_A)}{2}\ket{\psi}\nn\\
=&\frac{1}{\tan\theta}P^0_C X'_A\frac{(1-Z'_B)}{2}\ket{\psi}+\tan\theta P^0_C X'_A\frac{(1+Z'_B)}{2}.
\end{align}
Multiply by operator $(1+Z'_B)$ on both sides of \eqref{relation:Til3.1} such that 
\begin{align} 
P^0_C (1+Z'_B)X'_B\ket{\psi}
=&\frac{1}{\tan\theta}P^0_C X'_A(1+Z'_B)\frac{(1-Z'_B)}{2}\ket{\psi}\nn\\
&+\tan\theta P^0_C X'_A(1+Z'_B)\ket{\psi}\nn \\
=&\tan\theta P^0_C X'_A(1+Z'_A)\ket{\psi}\nn\\
=&P^0_C X'_B(1-Z'_B)\ket{\psi}
\end{align}
holds with \eqref{relation:Til1}, i.e.
\begin{align} \label{relation:Til4}
  P^0_CZ'_BX'_B\ket{\psi}=-P^0_C X'_B Z'_B\ket{\psi}.
\end{align}

Observation \eqref{ZAperXB} implies
\begin{align}
\langle P^0_C A_2(B_2-B_1)\rangle=0\Rightarrow P^0_C X'_A\ket{\psi} \perp P^0_C Z'_B\ket{\psi}
\end{align}
and combine the first relation \eqref{relation:Til1} from tilted Bell inequality, we have
\begin{align}\label{eq:P0CZaXa}
P^0_C X'_A\ket{\psi} \perp P^0_C Z'_A\ket{\psi}.
\end{align}
Then $Z_A\ket{\psi}$ in the subsapce of projector $P^0_C$ can be written as
\begin{align}\label{eq:PC0ZA}
P^0_C Z_A \ket{\psi}=\cos \omega P^0_C Z'_A \ket{\psi}+\sin \omega P^0_C X'_A \ket{\psi}
\end{align}
by equations \eqref{ZAXA} and \eqref{eq:P0CZaXa}. So, one can define the vector $X_A\ket{\psi}$ orthogonal to $Z_A\ket{\psi}$ as
\begin{align}\label{eq:XA}
 P^0_CX_A\ket{\psi}=\sin \omega P^0_C Z'_A \ket{\psi}- \cos \omega P^0_C  X'_A \ket{\psi}.
\end{align}
Since operators $Z'_A$ and $X'_A$ are hermitian, unitary
 and anti-commutation in the subspace of $P^0_C$ by \eqref{relation:Til1} and \eqref{relation:Til2}, 
 we get the anti-commutation relations
\begin{align}\label{eq:Anti-A}
P^0_C Z_AX_A\ket{\psi}=-P^0_CX_AZ_A\ket{\psi}.
\end{align}

Following the self-testing of maximally entangled
qubits from observation \cite{Wang}, maximal violation of the XOR game \eqref{eq:CHSH}
implies
\begin{align}\label{eq:2PC}
P^1_C Z'_A\ket{\psi}&=P^1_CZ'_B\ket{\psi},\nn\\
P^1_C X'_A\ket{\psi}&=P^1_CX'_B\ket{\psi}
\end{align}
and the anti-commutation relations
\begin{align}\label{eq:2PAnti1}
P^1_C Z'_AX'_A\ket{\psi}&=-P^1_CX'_AZ'_A\ket{\psi},\nn\\
P^1_C Z'_BX'_B\ket{\psi}&=-P^1_CX'_BZ'_A\ket{\psi}.
\end{align}

Observation \eqref{ZAperXB} implies
\begin{align}
\langle P^1_C A_2(B_2-B_1)\rangle=0\Rightarrow P^1_C X'_A\ket{\psi} \perp P^0_C Z'_B\ket{\psi}
\end{align}
and combine the first relation in \eqref{eq:2PC}, we have
\begin{align}\label{eq:PC1ZaXa}
P^1_C X'_A\ket{\psi} \perp P^1_C Z'_A\ket{\psi}.
\end{align}
Then $Z_A\ket{\psi}$  and  its orthogonal vector in the subsapce of projector $P^1_C$ can be written as
\begin{align}\label{eq:PC1ZA}
P^1_C Z_A\ket{\psi}&=\cos \omega P^1_C Z'_A \ket{\psi}+\sin \omega P^1_C X'_A \ket{\psi},\nn \\
P^1_CX_A \ket{\psi}&=\sin\omega P^1_C Z'_A \ket{\psi}-\cos\omega P^1_C X'_A \ket{\psi}
\end{align}
by equations \eqref{ZAXA} and \eqref{eq:PC1ZaXa}. 
Moreover, we can obtain the equivalence relations 
\begin{align}\label{eq:2PC2}
P^1_C Z_A\ket{\psi}&=P^1_CZ_B\ket{\psi},\nn\\
P^1_C X_A\ket{\psi}&=P^1_CX_B\ket{\psi}
\end{align}
and anti-commutation relations
\begin{align}\label{eq:2PAnti2}
P^1_C Z_AX_A\ket{\psi}&=-P^1_CX_AZ_A\ket{\psi},\nn\\
P^1_C Z_BX_B\ket{\psi}&=-P^1_CX_BZ_B\ket{\psi}
\end{align}
for party A in the subspace of $P^1_C$.

After some manipulations similar to party A, we can obtain
the relations between $Z_{B(C)}\ket{\psi}$, $X_{B(C)}\ket{\psi}$ and
$Z'_{B(C)}\ket{\psi}$, $X'_{B(C)}\ket{\psi}$ for party B (and C)
in the subspace of the projectors $P^0_C(P^0_A)$ and $P^1_C(P^1_A)$
\begin{align}\label{eq:ZAZB XAXB}
Z_{B(C)}\ket{\psi}&=\cos\omega Z'_{B(C)}\ket{\psi}+\sin\omega X'_{B(C)}\ket{\psi},\nn\\
X_{B}\ket{\psi}&=\cos\omega X'_{B}\ket{\psi}-\sin\omega Z'_{B}\ket{\psi},\nn\\
X_{C}\ket{\psi}&=\sin \omega Z'_{C}\ket{\psi}-\cos\omega X'_{C}\ket{\psi}.
\end{align}
Then the anti-commutation relations
\begin{align}\label{eq:Anti-C}
P^0_C Z_BX_B\ket{\psi}&=-P^0_CX_BZ_B\ket{\psi},\nn\\
P^1_C Z_BX_B\ket{\psi}&=-P^1_CX_BZ_B\ket{\psi},\nn\\
P^0_A Z_CX_C\ket{\psi}&=-P^0_CX_CZ_C\ket{\psi},\nn\\
P^1_A Z_CX_C\ket{\psi}&=-P^1_AX_CZ_C\ket{\psi}
\end{align}
in subspace hold.

Observations \eqref{Cor:XCXA} and \eqref{Cor:XB} imply that
\begin{subequations} 
\begin{align}\label{eq:canXC}
P^0_A P^0_B\ket{\psi}&=P^0_AP^0_BX_C\ket{\psi},\\
\label{eq:canXA}
P^0_BP^0_C\ket{\psi}&=P^0_BP^0_CX_A\ket{\psi},\\
\label{eq:canXB}
P^0_AP^0_C\ket{\psi}&=P^0_AP^0_CX_B\ket{\psi}.
\end{align}
\end{subequations}

Now the isometry can be constructed by $Z$, $X$ and $H$ per party as ref. 
\cite{Wu} as shown in Fig.1. 
The formula of $Z$ and $X$ 
are based on the measurement operators for each party, 
\begin{align}\label{Mea:Swap}
Z_A&=A_0,\;\;X_A=\sin\omega A_1-\cos\omega A_2;\nn \\
Z_B&=B_0,\;\;X_B=\frac{\cos(\omega-\mu)B_1}{\sin2\mu}-\frac{\cos(\omega+\mu)B_2}{\sin2\mu};\nn\\
Z_C&=C_0,\;\;X_C=\sin\omega C_1-\cos\omega C_2.
\end{align}
The output after the isometry can be written as
\begin{align}\label{eq:Iso}
\ket{\Psi'}=&\Phi\ket{\psi}_{ABC}\ket{000}_{A'B'C'}\nn\\
=&\frac{1}{8}[(I+Z_A)(I+Z_B)(I+Z_C)\ket{\psi}\ket{000}\nn\\
&+(I+Z_A)(I+Z_B)X_C(I-Z_C)\ket{\psi}\ket{001}\nn\\
&+(I+Z_A)X_B(I-Z_B)(I+Z_C)\ket{\psi}\ket{010}\nn\\
&+X_A(I-Z_A)(I+Z_B)(I+Z_C)\ket{\psi}\ket{100}\nn\\
&+X_A(I-Z_A)X_B(I-Z_B)X_C(I-Z_C)\ket{\psi}\ket{111}\nn\\
&+(I+Z_A)X_B(I-Z_B)X_C(I-Z_C)\ket{\psi}\ket{011}\nn\\
&+X_A(I-Z_A)(I+Z_B)X_C(I-Z_C)\ket{\psi}\ket{101}\nn\\
&+X_A(I-Z_A)X_B(I-Z_B)(I+Z_C)\ket{\psi}\ket{110}].
\end{align}

For the second term in \eqref{eq:Iso}, 
we could prove it is equal to $P^0_AP^0_BP^0_CX_C\ket{\psi}$ using \eqref{eq:Anti-C}.
Then, it can be replaced with $P^0_AP^0_BP^0_C\ket{\psi}$ because of \eqref{eq:canXC}.
The third and forth terms are similar.

For the fifth term in \eqref{eq:Iso}, 
one moves $X_A$ and $X_B$ to the right using \eqref{eq:2PAnti2}, which is in turn equal to  
$X_CP^1_CP^0_BX_BP^0_AX_A\ket{\psi}$. Then replace $X_B$ with $X_A$ 
by \eqref{eq:ZAZB XAXB} and this line becomes $P^0_AP^0_BX_CP^1_C\ket{\psi}$.
After moving $X_C$ to the right using \eqref{eq:Anti-C}, this term is equal to
$P^0_AP^0_BP^0_C\ket{\psi}$ from \eqref{eq:canXC}.

Remind that the last three terms in \eqref{eq:Iso} equal to zeros.

Therefore, all these properties of the operators deduced from the
measurement requirements will help to reduce the general
output \eqref{eq:Iso} to
\begin{align}\label{eq:outP}
\ket{\Psi'}=P^0_AP^0_BP^0_C\ket{\psi}(\ket{000}+\ket{001}+\ket{010}+\ket{100}+\ket{111})\nn.
\end{align}
This state can be normalized into
the form of $\ket{junk}_{ABC}\otimes\ket{\psi}_{A'B'C'}$, here $\ket{junk}=\sqrt{5}P^0_AP^0_BP^0_C\ket{\psi}$.

 Thus, we have proven that,
with these requirements \eqref{eq:Cor1}--\eqref{eq:CHSH} on the measurement
results indeed self-test the unknown state as target state \eqref{eq:target1}.

Further more, we also consider the robustness
for each observation in \eqref{eq:Cor1}--\eqref{eq:CHSH} 
has a deviation at most equal to $\epsilon$ around the perfect value. 
The robustness bound is given in the next section together (see Fig. 3).

\subsection{Robust self-testing of more general pure three-qubit states}

The previous work on self-testing of $W$ and $GHZ$ states
proved that both representatives
of the two inequivalent local operations and classical
communication classes of three-qubit \cite{DurW} can be self-tested.
The question then remains whether one can self-test every
pure three-qubit state. Here we explicitly shows how one can
self-test a large family of three-qubit states using swap method and
Navascu\'{e}s-Pironio-Ac\'{\i}n (NPA) hierarchy \cite{Navascues1}. 
The target state we consider is given as
\begin{equation}\label{eq:state}
\ket{\psi(\theta)}=\cos\theta\ket{W}+\sin\theta\ket{GHZ}
\end{equation}
here $\ket{W}=\frac{1}{\sqrt{3}}(\ket{001}+\ket{010}+\ket{100})$ and $\ket{GHZ}=\frac{1}{\sqrt{2}}(\ket{000}+\ket{111})$, $\theta\in[0,\frac{\pi}{2}]$ is a parameter .

\begin{result}
Alice and Bob each party performs two dichotomic measurements, 
Charlie performs three measurements with binary outcomes on an
unknown shared quantum state. The state $\ket{\psi_\theta}$ can be self-tested using the full statistics for $\theta\in[0,\frac{\pi}{2}]$. Moreover, the self-testing is robust.
\end{result}

We consider the $[2,2,3]$ scenario that Alice and Bob each performs two dichotomic measurements $Z$ and $X$ with binary outcomes as $\pm 1$, while Charlie performs three dichotomic measurements denoted as $Z$, $X$, and $D$ with outcomes. Suppose that the observed behavior exhibits the following two groups of full-body statistics
\begin{align}\label{eq:3PCor1}
&\braket{Z_iZ_jZ_k}=-\cos\theta^2,~~
\braket{Z_iX_jX_k}=\frac{2\cos^2\theta}{3}-\frac{\sin2\theta}{\sqrt{6}}, \nn\\
&\braket{Z_iZ_jX_k}=\frac{\sin2\theta}{\sqrt{6}},\;\;\;\;
\braket{X_iX_jX_k}=\sin\theta^2
\end{align}
up to permutations of Alice, Bob and Charlie, (i.e. $i,j,z\in\{A,B,C\}$, $i\neq j\neq z$) and
\begin{align}\label{eq:3PCor2}
&\braket{Z_i Z_j D_C}=-\frac{\cos^2\theta}{\sqrt{2}}+\frac{\sin2\theta}{2\sqrt{3}},\;\;
\braket{Z_i X_j D_C}=\frac{\sqrt{2}\cos^2\theta}{3},\nn\\
&\braket{X_i X_j D_C}=\frac{5\sqrt{2}-\sqrt{2}\cos2\theta-2\sqrt{3}\sin2\theta}{12}
\end{align}
up to permutations of Alice and Bob (i.e. $i,j\in\{A,B\}$,  $i\neq j$).

These are
the statistics that one would obtain for the $\ket{\psi}$ for $\theta\in[0,\frac{\pi}{2}]$ if
$Z=\sigma_z$, $X=\sigma_x$ and $D=\frac{\sigma_x+\sigma_z}{\sqrt{2}}$.

We consider the same isometry as sec.\ref{sec:3.1} as shown in
Fig. 1. The isometry can be re-written as a swap operator $U=S_{AA'}\otimes
 S_{BB'}\otimes S_{CC'}$ with $S_{AA'} =W_{AA'}V_{AA'}W_{AA'}$ and
\begin{align}
W_{AA'}=&I_A\otimes\ket{0}\bra{0}+X_A\otimes\ket{1}\bra{1},\nn\\
V_{AA'}=&\frac{I_A+Z_A}{2}\otimes I+\frac{I_A-Z_A}{2}\otimes\sigma_x
\end{align}
and the same for $S_{BB'}$ and $S_{CC'}$. After this isometry, the trusted auxiliary systems
will be left in the state
\begin{align}
\rho_{swap}=&tr_{ABC}[U\rho_{ABC}\otimes\ket{000}\bra{000}_{A'B'C'}U^\dagger]\nn\\
=&\sum C_{ijklst}\ket{i}\bra{j}\otimes\ket{k}\bra{l}\otimes\ket{s}\bra{t}
\end{align}
where
\begin{align}
C_{ijklst}=\frac{1}{64}tr_{ABC}[M^A_{j,i}\otimes M^B_{l,k}\otimes M^C_{t,s}\rho_{ABC}]
\end{align}
and $M^A_{j,i}=(I+Z_A)^{j+1}(X_A-X_AZ_A)^j(I+Z_A)^{i+1}(X_A-X_AZ_A)^i$ for $i,j\in\{0,1\}$,
$M^B_{l,k}$ and $M^C_{t,s}$ are analogous.

Finally, the closeness of the unknown resource to the target state
can be then captured by the fidelity
\begin{align}f=\bra{\psi}\rho_{swap}\ket{\psi}
\end{align}
 as a linear function of $\theta$, observed behavior and some non-observable correlations which 
 involve different measurements on the same party, 
such as $\bra{\psi}M^a_x M^{a'}_{x'}\ket{\psi}$ with $x\neq x'$ which are left as 
 variables.
The terms in fidelity that are not determined should be compatible
with a quantum realisation. As well known, this requirement can't be formulated
as an efficient constraint, but it can be relaxed to a family of semi-definite constraints\cite{Navascues1}. 
Since the objective function $f$ is linear, the optimisation can then
be cast as a semi-definite programming (SDP) \cite{Yang2,Bancal}:
\begin{align}
\label{eq:MSDP}
\min \quad & f=\bra{\psi}\rho_{swap}\ket{\psi} \nn\\
\text{s.t.}\quad  &\Gamma \geq 0, \\
& \text{equations}~\eqref{eq:3PCor1} \text{ and } \eqref{eq:3PCor2}, \nn
\end{align}
where $\Gamma$ is a matrix with NPA
hierarchy characterization of the quantum behaviors.
This moment matrix  corresponding to $q$-local level 1 (which
includes any products with at most one operator per party) has size $74\times74$ 
and is augmented by necessary terms(like $\langle Z_AX_AZ_A\rangle$,
$\langle Z_BX_B Z_C X_C\rangle$, $\langle Z_C X_C Z_AX_AZ_A\rangle$ and so on) to express all the average values $\langle \cdot\rangle$ that
appear in the expression of fidelity. For all $\theta\in[0,\frac{\pi}{2}]$, the SDP returns $f>99.96\%$ (Fig.2). We believe that the deviation from 1 is due to
the limitation of the SDP relaxation.
\begin{figure}[ht]\label{fig:rob3P}
\centering
\includegraphics[scale=0.43]{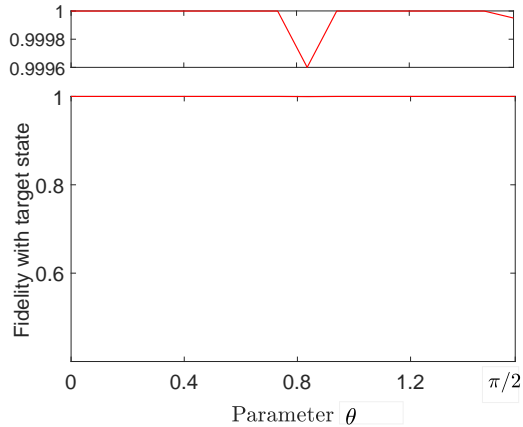}
\caption{Red line represents the lower bound on the fidelity obtained with varying parameter $\theta$ from 0 to $\frac{\pi}{2}$.}
\end{figure}

Interesting as the above result is in itself, it relies on
observing the measurement statistics in \eqref{eq:3PCor1} and \eqref{eq:3PCor2}
exactly, which is not possible due to inevitable experimental
uncertainties. To investigate the robustness of self-testing induced by these statistics,
we shall consider mixing them with white noise, that is
by multiplying each term by $(1-\epsilon)$ and $\epsilon$ represents the deviation of the observed behavior
from the ideal values.
As examples, we plot four special values for $\theta =\{0, \frac{\pi}{4}, \frac{\pi}{2}, \arccos\sqrt{\frac{3}{5}}\}$ which correspond to $\ket{W}$, $\frac{\ket{W}+\ket{GHZ}}{\sqrt{2}}$,
$\ket{GHZ}$ and superposition state investigated in sec.\ref{sec:3.1}  $$\frac{1}{\sqrt{5}}(\ket{000}+\ket{001}+\ket{010}+\ket{100}+\ket{111}),$$
respectively (Fig.3).

\begin{figure}[ht]\label{fig1}
\centering
\includegraphics[scale=0.45]{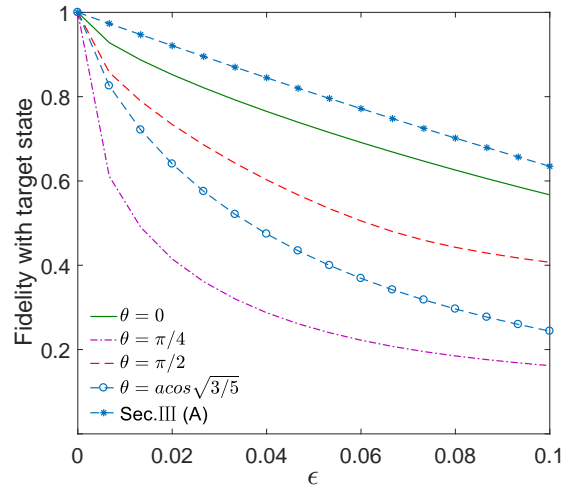}
\caption{(Color Online) The plot of lower bounds on the fidelity
between the measured state and the target state.
$\epsilon$ represents the deviation of the observed behavior
from the ideal values. }
\end{figure}

Note that even though calculation of deducing the expression of fidelity does not contain any
moment involving the measurement $D$, its appearance in the
matrix $\Gamma$ makes it necessary and useful to bound the fidelity.

In particular, the robustness for self-testing the specific state analysed in sec.\ref{sec:3.1} can be also obtained by SDP. 
Let us first look at the ideal quantum realization to design our swap circuit. 
 To construct the swap circuit, set $Z_A=A_0$, $Z_B=B_0$ and $Z_C=C_0$, 
 and we'd rather need $\sigma_x$ per party, 
which in the ideal case should have the forms as \eqref{Mea:Swap}.
However, written with the unknown measurement operators, this expression may not define a unitary operator. A method to circumvent this obstacle has been presented in previous works~\cite{Wang,Bancal}: one defines a fourth dichotomic operator $\widetilde{X}_i$ for $i\in\{A,B,C\}$ such that 
\begin{align}
\widetilde{X}_A(\sin{\omega}A_1-\cos\omega A_2)&\geq0,\nn\\
\widetilde{X}_B\frac{\cos(\omega-\mu)B_1-\cos(\omega+\mu)B_2}{\sin(2\mu)}&\geq 0,\nn\\
\widetilde{X}_C(\sin{\omega}C_1-\cos\omega C_2)&\geq0.
\end{align}
Since these equations are not SDP constraints, one relaxes each equation to the positivity of a ``localizing matrix''.  

In order to make the lower bound tight, we add a localizing matrix for $\sigma_z$ each party. 
Then run the SDP using NPA matrix size $163\times 163$ and augmented by six localizing matrices (two per party), 
to minimize the fidelity with the target state for each observation in \eqref{eq:Cor1}--\eqref{eq:CHSH} with a deviation $\epsilon$ from the perfect value. The result is summarized in Fig.3.


\section{Conclusion}
We proposed self-testing schemes for a large family of symmetric three-qubit states. The target states we mainly focused on is the superposition of $W$ state and $GHZ$ state due to the simple form and their wide applications in quantum information tasks. We provided two different approaches applying to these states.

For the special case where the state has equal coefficients of the canonical basis, our approach is constructed by combining bipartite self-testing schemes. Through projecting the state  onto  two kinds of subsystem entangled states by one party's measurement $Z$,  the remaining parties can reach the self-testing criteria for these bipartite entangled states simultaneously using the same measurements settings. The bound is robust against the inevitable experimental errors. 

For the general case, we demonstrated that these states can be self-tested using fixed measurements numerically. Here in our work, only the simplest Pauli measurements in $[2,2,3]$ scenario are used. This is quite helpful in the experiments. It would be of interest to study whether this result could be generalized to generic multipartite states. Previous work on multipartite states are usually realized by constructing tailored Bell inequalities for the target states. The complexity of self-testing multipartite states would decrease significantly if our approach holds. A comprehensive study of these questions remains open for other states and scenarios.

%


\section{Appendix}
This appendix provides the details of the 
transformational relations between the Pauli matrices and measurements operators 
by Schmidt decomposition.

Rewrite the state 
   $$\ket{\varphi_0}:=\frac{1}{\sqrt{3}}(\ket{00}+\ket{01}+\ket{10}).$$
So we can denote the coefficient matrix as
\begin{align}
C=\frac{1}{\sqrt{3}}{\left[ \begin{array}{cc}
1 & 1 \\
1 &  0 
\end{array} 
\right ]}
\end{align}
which has Schmidt decomposition $C=USV$, where 
\begin{align}
S={\left[ \begin{array}{cc}
\sqrt{\frac{1}{6}(3+\sqrt{5})} & 0 \\
0 &  \sqrt{\frac{1}{6}(3-\sqrt{5})} 
\end{array} 
\right ]}
\end{align}
is diagonal, $U$ and $V$ are unitary matrices:
\begin{align}
U={\left[ \begin{array}{cc}
\frac{3+\sqrt{5}}{2\sqrt{5+2\sqrt{5}}} & \frac{3-\sqrt{5}}{2\sqrt{5-2\sqrt{5}}}  \\
\frac{1+\sqrt{5}}{2\sqrt{5+2\sqrt{5}}}  & \frac{1-\sqrt{5}}{2\sqrt{5-2\sqrt{5}}}  
\end{array} 
\right ]},
\;\;
V={\left[ \begin{array}{cc}
\frac{1+\sqrt{5}}{2\sqrt{5+\sqrt{5}}} & \frac{1-\sqrt{5}}{\sqrt{10-2\sqrt{5}}}  \\
\sqrt{\frac{2}{5+\sqrt{5}}}  & \frac{5+\sqrt{5}}{10} 
\end{array} 
\right ]}.
\end{align}
 Define $\ket{j'_A}\equiv \sum\limits_{j}U_{ji}\ket{j}$, 
$\ket{k'_B}\equiv\sum\limits_{k}V_{ik}\ket{k}$ and $\lambda_i=S_{ii}$, 
for $i,j,k\in\{0,1\}$. Obviously, $\{\ket{j'}\}_A$ and $\{\ket{k'}\}_B$
are two groups new standard orthogonal basis. Then 
the state $\ket{\varphi_0}$ can be written as
\begin{align}
\ket{\varphi_0}=&\sum\limits_{ijk}U_{ji}S_{ii}V_{ik}\ket{j}_A\ket{k}_B\nn\\
=&\cos\beta\ket{0'_A0'_B}+\sin\beta\ket{1'_A1'_B},
\end{align}
here $\cos \beta =\sqrt{\frac{3+\sqrt{5}}{6}}$ and $\sin\beta=\sqrt{\frac{3-\sqrt{5}}{6}}$.

Now, consider the relation between operator $Z'$ and $X'$ 
in the new basis and $\sigma_z$ and $\sigma_x$
for part A,
\begin{align}
    Z'_A=&\ket{0'}\bra{0'}_A-\ket{1'}\bra{1'}_A\nn\\
    =&\sum_{j_1,j_2\in\{0,1\}^2}U_{j_10}U_{j_20}\ket{j_1}\bra{j_2}
    -\sum_{j_1,j_2\in\{0,1\}^2}U_{j_11}U_{j_21}\ket{j_1}\bra{j_2}\nn\\
    =&\frac{1}{\sqrt{5}}(\ket{0}\bra{0}-\ket{1}\bra{1})+\frac{2}{\sqrt{5}}(\ket{0}\bra{1}+\ket{1}\bra{0})\nn\\
    =&\cos\omega \sigma_z+\sin\omega \sigma_x,
\end{align}
here $\tan\omega=2$. It is easy to get $X'_A=\sin\omega\sigma_z-\cos\omega\sigma_x$. 
This is also hold for part C. Similarly, we have
\begin{align}
    Z'_B=&\ket{0'}\bra{0'}_B-\ket{1'}\bra{1'}_B\nn\\
    =&\sum_{k_1,k_2\in\{0,1\}^2}V_{0k_1}V_{0k_2}\ket{k_1}\bra{k_2}
    -\sum_{k_1,k_2\in\{0,1\}^2}V_{1k_1}V_{1k_2}\ket{k_1}\bra{k_2}\nn\\
    =&\frac{1}{\sqrt{5}}(\ket{0}\bra{0}-\ket{1}\bra{1})-\frac{2}{\sqrt{5}}(\ket{0}\bra{1}+\ket{1}\bra{0})\nn\\
    =&\cos\omega \sigma_z-\sin\omega \sigma_x
\end{align}
and $X'_B=\sin\omega \sigma_z+\cos\omega\sigma_x$.

Hence, if Alice and Bob each performs optimal operators as \eqref{M Tilted} using the new basis, 
then the measurements can be transformed into Pauli matrices
\begin{equation}
\begin{cases}
\sigma_z=\cos\omega A_1+\sin\omega A_2,\\
\sigma_x=\sin\omega A_1-\cos\omega A_2;
\end{cases}
\hspace{1em}
\begin{cases}
\sigma_z=\frac{\sin(\omega+\mu)B_2-\sin(\omega-\mu)B_1}{\sin2\mu},\\
\sigma_x=\frac{\cos(\omega-\mu)B_1-\cos(\omega+\mu)B_2}{\sin2\mu}.
\end{cases}
\end{equation}
Charlie is analogue to Alice.
\section*{Acknowledgment}

We would like to thank Valerio Scarani, Huangjun Zhu and Yu Cai for useful discussions. 
This work is funded by National Nature Science Foundation of China 
(Grant Nos. 61671082, 61572081, 61672110, 11875110).


\end{document}